\documentstyle[aps,psfig,12pt]{revtex}
\def\step{\\[-1.5ex]}

\def\beq{\begin{equation}}
\def\eeq{\end{equation}}

\def\bea{\arraycolsep .1em \begin{eqnarray}}
\def\eea{\end{eqnarray}}
\def\Tr{{\rm Tr}}

\def\grgl{\:\hbox to -0.2pt{\lower2.5pt\hbox{$\sim$}\hss}{\raise3pt\hbox{$>$}}\:}
\def\klgl{\:\hbox to -0.2pt{\lower2.5pt\hbox{$\sim$}\hss}{\raise3pt\hbox{$<$}}\:}
\def \lta {\mathrel{\vcenter
     {\hbox{$<$}\nointerlineskip\hbox{$\sim$}}}}
\def \gta {\mathrel{\vcenter
     {\hbox{$>$}\nointerlineskip\hbox{$\sim$}}}}

\let\Ga=\Gamma

\def\eq#1{(\ref{#1})}
\def\Eq#1{(\ref{#1})}
\def\Es#1{(\ref{#1})}

\def\s0#1#2{\mbox{\small{$ \frac{#1}{#2} $}}}
\def\0#1#2{\frac{#1}{#2}}

\makeatletter

\renewenvironment{thebibliography}[1]
         {\frenchspacing\small
          \begin{list}{[\arabic{enumi}]}
         {\usecounter{enumi}\parsep=2pt\topsep 0pt
         \settowidth{\labelwidth}{[#1]}
         \leftmargin=\labelwidth\advance\leftmargin\labelsep
         \rightmargin=0pt\itemsep=0pt\sloppy}}{\end{list}}

\makeatother

\begin{document}
\begin{center}

\thispagestyle{empty}

{\normalsize\begin{flushright} 
CERN-TH-2001-321\\[12ex] 
\end{flushright}}

\mbox{\large \bf Derivative expansion and renormalisation group flows}
\\[6ex]

{Daniel F. Litim}
\\[2ex]
{\it CERN Theory Division, CH -- 1211 Geneva 23.}
\\
{\footnotesize (Daniel.Litim@cern.ch)}
\\[10ex]
 
{\small \bf Abstract}\\[2ex]
\begin{minipage}{14cm}{\small 
    We study the convergence of the derivative expansion for flow
    equations. The convergence strongly depends on the choice for the
    infrared regularisation. Based on the structure of the flow, we
    explain why optimised regulators lead to better physical
    predictions. This is applied to $O(N)$-symmetric real scalar field
    theories in $3d$, where critical exponents are computed for all
    $N$. In comparison to the sharp cut-off regulator, an optimised
    flow improves the leading order result up to $10\%$.  An analogous
    reasoning is employed for a proper time renormalisation group. We
    compare our results with those obtained by other methods. }
\end{minipage}
\end{center}

\newpage
\pagestyle{plain}
\setcounter{page}{1}
\noindent 
{\bf 1. Introduction}\\[-1ex]

The recent years have witnessed an important progress in the
development of non-perturbative methods that provide a reliable
description of systems with large effective couplings or divergent
correlation lengths, problems which are difficult to handle, if at
all, within standard perturbation theory. A particularly efficient
method is known as the exact renormalisation group (ERG)
\cite{Polchinski,continuum,Ellwanger:1994mw,Morris:1994qb} (for recent
reviews, see \cite{Bagnuls:2000ae,Review} for scalar and
\cite{Litim:1998nf} for gauge theories), which grew out of the idea of
coarse-graining quantum fields. In its modern form, the ERG flow for
an effective action $\Gamma_k$ for bosonic fields is given by the
simple one-loop expression
\beq\label{ERG}
\partial_t \Gamma_k = 
\s012 {\rm Tr} \left(\Gamma^{(2)}_k+R_k\right)^{-1}\partial_t R_k\,.
\eeq
Here, $\Gamma^{(2)}_k$ denotes the second functional derivative of the
effective action, $t=\ln k$ is the logarithmic scale parameter,
and $R_k(q^2)$ is an infrared (IR) regulator at the momentum scale
$k$.  The regulator fulfils a few constraints (given below) to ensure
that \Eq{ERG} interpolates between a microscopic initial action
$\Gamma_\Lambda$ at the ultraviolet (UV) scale $k=\Lambda$, and the
full quantum effective action $\Gamma$ in the infrared limit
$k=0$.\step

An application of the ERG formalism requires some approximations. A
commonly used systematic approximation scheme is the derivative
expansion of effective actions \cite{Golner:1986}.  However, little is
known about its convergence, because there is {\it a priori} no small
parameter associated to it.  Indeed, for the ERG flow \eq{ERG}, the
derivative expansion implies an expansion of $\Gamma_k^{(2)}$ inside
the momentum trace in powers of $q/k$, where $q$ is the loop momenta
and $k$ the infra-red scale. The validity of such a procedure, and,
consequently, the convergence of the derivative expansion, are {\it
  unavoidably} entangled with the particular choice for the IR
regulator, since the regulator ensures that the momentum trace is
peaked for momenta $q^2\approx k^2$.  For the computation of
$\beta$-functions, this has been studied in \cite{Morris:1999ba}. More
generally, and similar to perturbative QCD or truncations of Schwinger
Dyson equations, approximate solutions of \eq{ERG} depend spuriously
on the IR regularisation
\cite{Ball:1995ji,Litim:1997nw,Freire:2001sx,Liao:2000sh,Litim:2000ci,Litim:2001up,Litim:2001fd}.
A deeper understanding of the scheme dependence, and its link to the
stability and convergence of the ERG flow, has been established
recently \cite{Litim:2000ci,Litim:2001up,Litim:2001fd}. These results
are at the basis for reliable physical predictions based on the ERG
formalism.  \step

In this Letter, we study the convergence of the derivative expansion
for $3d$ $O(N)$-symmetric scalar theories. We compute universal
critical exponents for all $N$ and different regularisations.  Based
on conceptual arguments, we explain why specific regulators are
expected to provide better physical predictions already to leading
order in the derivative expansion. A deeper link between the
convergence and the optimisation of ERG flows is established.  While
most of our considerations are based on ERG flows, we also discuss the
convergence for specific proper time flows.  A detailed comparison
with results from other methods and higher orders in the derivative
expansion is also given.
\\[3ex]

\noindent 
{\bf 2. Derivative expansion}\\[-1ex]

In the context of flow equations, the derivative expansion is based on
the assumption that higher derivative operators lead only to small
corrections compared to lower order ones, which does not imply that
the UV and IR degrees of freedom are the same. Hence, the anomalous
dimension $\eta$ of the quantum fields should be small. Some physical
systems are known where these assumptions are realised. Consider the
universality class of $O(N)$-symmetric real scalar theories in $d=3$
dimensions, where $N=0$ describes the physics of entangled polymers;
$N=1$ the Ising model (water); $N=2$ the XY model (He${}^4$); $N=3$
the Heisenberg model (ferromagnets).  The case $N=4$ is expected to
describe the phase transition of QCD with two massless quark flavours.
\step

At the Wilson-Fisher fixed point the critical properties of
$O(N)$-symmetric real scalar theories are characterised by universal
critical exponents: $\nu_{\rm phys}$, given by the inverse of the
negative eigenvalue of the stability matrix at criticality, and
$\eta_{\rm phys}$, the anomalous dimension.  It is known from
experiment that $\eta_{\rm phys}$ is at most of the order of a few
percent.  Hence, it is believed that the derivative expansion is a
good approximation for a reliable computation of universal critical
exponents.  Within the derivative expansion, the physical critical
exponents at the scaling solution are computed as the series
\bea\label{nu-expansion}
\nu_{\rm phys} =&\nu_{0}({\rm RS})&  + \nu_{1}({\rm RS})
                                     + \nu_{2}({\rm RS})+\cdots\\
\label{eta-expansion}
\eta_{\rm phys}=&0&                  +\eta_{1}({\rm RS})
                                     +\eta_{2}({\rm RS})+\cdots
\eea
Here, the index corresponds to the order of the derivative expansion.
To leading order, $\eta_{0}({\rm RS})\equiv 0$. Notice that every
single order in the expansion --- due to the approximations employed
--- depends on the regulator scheme.  The independence of physical
observables on the regulator scheme (RS) can only be guaranteed in the
limit where {\it all} operators of the effective action are retained
during the flow.  In turn, the physical values $\nu_{\rm phys}$ and
$\eta_{\rm phys}$ are independent of the precise form of the infrared
regulator. Hence, the infinite sum on the right-hand side adds up in a
way such that the physical values are scheme independent. The
convergence of the expansion \eq{nu-expansion} is best if a regulator
is found such that the main physical information is contained in a few
leading order terms. \step

For specific values of $N$, the derivative expansion becomes exact
(and hence independent on the regulator). In the large-$N$ limit, the
universal critical exponents are
\bea
\nu_{\rm  phys}     &=& 1 + {\cal O}(1/N)\label{nu-largeN} \\
\omega_{\rm  phys}  &=& 1 + {\cal O}(1/N)\label{omega-largeN} \\
\eta_{\rm phys}     &=& 0 + {\cal O}(1/N)\,.
\eea
All subleading universal eigenvalues of the stability matrix at
criticality are given as $\omega_n=2n-1+{\cal O}(1/N)$ for $n\ge 1$
($\omega\equiv \omega_1)$, and all ${\cal O}(1/N)$ coefficients are RS
dependent. Stated differently, the large-$N$ limit is so strong that
all scheme dependence of universal quantities is suppressed as ${\cal
  O}(1/N)$.  For the case $N=-2$, it is known that
\cite{BT:1973,F:1973}
\bea
\nu_{\rm  phys}  &=& \s012+ {\cal O}(N+2) \label{nu-2} \\
\eta_{\rm phys}  &=&   0  + {\cal O}(N+2)\,,
\eea
for all regulators. All RS dependent corrections are supressed.  In
contrast to the large-$N$ limit, we find that the subleading
eigenvalues $\lambda_n$ are {not} universal (see Fig.~2 below).
\\[3ex]

\noindent 
{\bf 3. Renormalisation group flows}\\[-1ex]

Next we use \Eq{ERG} to compute the leading order of \Eq{nu-expansion}
for all $N$. The flow trajectory of \Eq{ERG} in the space of all
action functionals depends on the IR regulator function $R_k$. $R_k$
obeys a few restrictions, which ensure that the flow equation is
well-defined, thereby interpolating between an initial action in the
UV and the full quantum effective action in the IR. We require that
\bea
\label{I}
\lim_{q^2/k^2\to 0}R_k(q^2)>0\,,     \\
\label{II}
\lim_{k^2/q^2\to 0}R_k(q^2)\to 0\,,  \\
\label{III}
\lim_{k\to\Lambda}R_k(q^2)\to \infty\,.
\eea
Equation \eq{I} ensures that the effective propagator at vanishing
field remains finite in the infrared limit $q^2\to 0$, and no infrared
divergences are encountered in the presence of massless modes.
Equation \eq{II} guarantees that the regulator function is removed in
the physical limit, where $\Ga\equiv\lim_{k\to 0}\Ga_k$.  Equation
\eq{III} ensures that $\Ga_k$ approaches the microscopic action
$S=\lim_{k\to \Lambda}\Ga_k$ in the UV limit $k\to \Lambda$.  We put
$\Lambda=\infty$, and introduce for later convenience the
dimensionless regulator function $r(y)$ as 
\beq\label{r}
R_k=Z_k\cdot q^2\, r(q^2/k^2)\,.
\eeq
Here, we have also introduced a wave function renormalisation factor
$Z_k$ into the regulator. We set $Z\equiv 1$ to leading order in the
derivative expansion. A few explicit regulators are
\bea
\label{ropt}
r_{\rm opt}(y)     & = & (\s01y -1)\theta(1-y)      \\
\label{rpower}
r_{\rm power}(y|b) & = & y^{-b}                     \\
\label{rsharp}
r_{\rm sharp} (y)  & = & 1/\theta(1-y)-1            \\
\label{rexp}
r_{\rm exp}(y|b)   & = & 1/(\exp c y^b -1)            \\
\label{rmix}
r_{\rm mix}(y|b)   & = & \exp[-b(y^{1/2}-y^{-1/2})] \,,
\eea
with $b\ge 1$ (and $c=\ln 2$). The optimised regulator \Eq{ropt} has
been introduced in Ref.~\cite{Litim:2001up}.  It leads to better
stability and convergence properties of ERG flows
\cite{Litim:2000ci,Litim:2001fd}.  The class of power-like regulators
\Eq{rpower} for $b\ge 1$ is often used for analytical and numerical
considerations \cite{Morris:1994ie}.  However, for large momenta, the
regulator decays only as a power law, as does $\partial_t R_{\rm
  power}$. This may lead to an insufficiency in the integrating out of
momentum variables.  Eq.~\eq{rsharp} describes the sharp cut-off
regulator. It corresponds to the limit $b\to\infty$ of both
\Eq{rpower} and \Eq{rexp} \cite{Tetradis:1996br}.  The exponential
regulator \Eq{rexp}, which is also expected to have good convergence
properties due to the exponential suppression of large momenta, is
often used for numerical studies
(e.g.~\cite{Review,VonGersdorff:2000kp}).  Finally, the class of mixed
exponential regulator $r_{\rm mix}$ has been employed in
\cite{Liao:2000sh}.  \step

With the main definitions at hand, we turn to the ERG flow for
$O(N)$-symmetric scalar theories to leading order in the derivative
expansion --- the local potential approximation. We are lead to the
Ansatz
\beq
\label{AnsatzGamma}
\Gamma_k=\int d^dx \left(U_k(\bar\rho) 
        + \012 Z_k(\bar\rho) \partial_\mu \phi^a\partial_\mu \phi_a
        + \014 Y_k(\bar\rho) \partial_\mu \bar\rho\partial_\mu \bar\rho
        +{\cal O}(\partial^4)
\right)
\eeq
for the effective action $\Gamma_k$, with
$\bar\rho=\s012\phi^a\phi_a$. For $N\neq 1$, there are two independent
wave function factors $Z_k$ and $Y_k$ beyond the leading order in this
expansion. To leading order in the derivative expansion, the flow
equation \eq{ERG} reduces to a flow $\partial_t U_k$ for the effective
potential.  Inserting \Eq{AnsatzGamma} into \Eq{ERG}, setting $Z\equiv
Y\equiv 1$, and rewriting the flow for the effective potential in
dimensionless variables $u=U_k/k^d$ and $\rho=\bar\rho/k^{d-2}$, we
find
\beq\label{GeneralFlow}
\partial_t u+du-(d-2) \rho u'
=2v_d(N-1)\ell(u')+2v_d \ell(2\rho u'')\,
\eeq
and $v^{-1}_d=2^{d+1}\pi^{d/2}\Gamma(\s0d2)$. Here, we have also
performed the angle integration of the momentum trace. The functions
$\ell(\omega)$ are given by
\beq\label{Id}
\ell(\omega)
=\s012\int^\infty_0dy y^{d/2}
\0{\partial_t r(y)}{y(1+r)+\omega}\,
\eeq
where the integration over $y\equiv q^2/k^2$ corresponds to the
remaining integral over the size of the loop momenta, and $\partial_t
r(y) = -2 y r'(y)$. All non-trivial information regarding the
renormalisation flow and the RS are encoded in \Eq{Id}. In turn, all
terms on the left-hand side of \Eq{GeneralFlow} do not depend on the
RS. They simply display the intrinsic scaling of the variables which
we have chosen for our parametrisation of the flow.  For the
regulators $r_{\rm opt}(y)$, $r_{\rm quart}(y)\equiv r_{\rm
  power}(y|2)$ and $r_{\rm sharp}(y)$, the flows \Eq{Id} can be
computed analytically. We find
\bea
\ell_{\rm opt}(\omega)  &= & \s02d\, (1+\omega)^{-1}\label{l-opt}\\
\ell_{\rm quart}(\omega)&= & {\pi}\, ({2+\omega})^{-1/2}\label{l-quart}\\
\ell_{\rm sharp}(\omega)&= & -\ln({1+\omega})\label{l-sharp}
\eea
While \Es{l-opt} and \eq{l-sharp} hold for any dimension, \Eq{l-quart}
holds only for $d=3$. Notice that the functions $\ell(\omega)$ differ
mainly in their asymptotic decay for large arguments.
\\[3ex]

\noindent 
{\bf 4. Convergence}\\[-1ex]

Now we address the convergence of \eq{nu-expansion}, which is expected
to be best if $|\nu_0({\rm RS})| \gg |\nu_1({\rm RS})| \gg \cdots\,,$
as this would imply that the main physical information is contained
within a few leading order terms of the expansion. Hence, it is
mandatory, first, to understand the range within which $\nu_0({\rm
  RS})$ varies as a function of the scheme \cite{Litim:CritExp}, and,
second, to discriminate those regulators for which $|\nu_1({\rm
  RS})/\nu_0({\rm RS})|$ is smallest.  In
Refs.~\cite{Litim:2000ci,Litim:2001up,Litim:2001fd}, we have argued
that specific such regulators --- based on an optimisation of the flow
--- are indeed available.  Consider the flow \eq{GeneralFlow},
parametrised in terms of the scheme dependent functions \Eq{Id}, in
the region of small (field) amplitudes.  With small amplitudes, we
denote those regions in field space where $|\omega|\klgl {\cal O}(1)$,
while $\omega\equiv u'$ or $u'+2\rho u''$ are the amplitudes entering
the functions \Eq{Id}. For example, a polynomial approximation of the
scaling potential about the local minimum is viable for a
determination of universal critical exponents \cite{Aoki:1998um}, with
the exception of flows based on problematic regulators like the sharp
cut off \cite{Morris:1994ki,Litim:2000ci}. Note that the expansion is
performed only for the presentation of our line of reasoning.  It is
certainly not needed for solving the equations.  Expanding the flow
\eq{Id} in powers of its argument, we find
\beq
\label{ell-w}
\ell(\omega)=\sum_{n=0}^\infty a_n (-\omega)^n
\eeq
with expansion coefficients ($y\equiv q^2/k^2$) 
\beq
\label{an}
a_n =\int^\infty_0 dy \0{-y^{1+d/2}\,r'(y)}{[y(1+r)]^{n+1}}\,,
\eeq
and all $a_n\ge 0$. For small and fixed $\omega$, only a few
coefficients $a_n$ are required for a reliable approximation of the
flow. Higher order corrections are smallest if $a_n/a_{n+1}$ is
smallest. In the limit $n\to\infty$, this ratio becomes the radius of
convergence of amplitude expansions,
\beq\label{C}
C=\lim_{n\to\infty}\left( a_n/a_{n+1} \right)
\eeq
In Ref.~\cite{Litim:2000ci}, it has been shown that the limit \Eq{C}
is given by
\beq\label{Cexpl}
C=\min_{y\ge 0}\ y(1+r(y))\,.
\eeq
The result is easily understood: the function $y(1+r(y))$, the
regularised dimensionless inverse propagator at vanishing field,
displays a ``gap'' due to the regularisation.  Consequently, in the
limit $n\to\infty$, the integrand in \Eq{an} is suppressed the least
at the minimum of $y(1+r)$, whence \Eq{Cexpl}.  Let us normalise all
regulators by the requirement $r(\s012)=1$, in order to compare their
respective radii of convergence. (The normalisation differs from the
one used in Ref.~\cite{Litim:2000ci}.)  From \Eq{Cexpl} and the
normalisation condition, we obtain $C_{{}_{\rm OPT}}\equiv\max_{\rm
  (RS)}C=1$. The gap is bounded from above. We refer to regulators
with $C_{{}_{\rm OPT}}=1$ as optimised. The extremisation of
\Eq{Cexpl} is closely linked to a minimum sensitivity condition
\cite{Litim:2001fd}.\step

In Ref.~\cite{Litim:2000ci}, we have computed the radii of convergence
for different classes of regulators. For \Eq{ropt}, we found $C_{\rm
  opt}=1$. For the sharp cut off, the result is $C_{\rm sharp}/C_{\rm
  opt}=\s012<1$. In turn, for the quartic regulator, we have $C_{\rm
  quart}/C_{\rm opt}=1$.  Therefore, we expect better convergence
properties for flows based on $r_{\rm quart}$ and $r_{\rm opt}$ than
those based on $r_{\rm sharp}$.  A last comment concerns all
regulators with $C=1$. While their asymptotic limits \Eq{C} are the
same, their approach to \Eq{C} still depends on the RS.  The specific
regulator \Eq{ropt} is distinguished because the limit \Eq{C} is
attained for {\it all} values of $n$, and not only asymptotically.
Therefore, we expect better results for flows based on
$r_{\rm opt}$ in \Eq{ropt}.\\[3ex]

\noindent 
{\bf 5. Anomalous dimension}\\[-1ex]

Next, we consider higher order corrections to \eq{nu-expansion}
implied through the anomalous dimension $\eta=-\partial_t \ln Z$.
Here, $Z^{1/2}$ stands for the wave function renormalsiation of the
field. These corrections appear at second order in the derivative
expansion. Once $\eta\neq 0$, the renormalised dimensionless field
variable is $\rho=\s012 Z_k\cdot \phi^a\phi_a/k^{d-2}$.  Consequently,
the term $(d-2)\rho u'$ on the left-hand side of \Eq{GeneralFlow}
changes into $(d-2-\eta)\rho u'$.  We also introduce $Z$ into the
regulator, as done in \Eq{r}.  Therefore, the scale derivative
$\partial_t r$ in \Eq{Id} depends on $\eta$ as $\partial_t r=-2yr'
-\eta\,r$.  The new function $\ell(\omega,\eta)$ is Taylor expanded as
\beq
\label{ell-w-eta}
\ell(\omega,\eta)=\sum_{n=0}^\infty 
\left(a_{n} -\eta \, b_{n}\right)(-\omega)^n\,,
\eeq
where $a_n$ are given by \Eq{an}, and the higher order expansion
coefficients $b_n\ge 0$ are
\beq
\label{bn}
b_n=\s012\int_0^\infty dy \0{y^{d/2}\,r(y)}{[y(1+r)]^{n+1}}\,.
\eeq
The back coupling of higher order operators --- parametrised by $\eta$
--- is proportional to the ratio $b_n/a_n$.  Consider this ratio in the
limit of large $n$. We find
\beq\label{R}
\lim_{n\to\infty}\left( \0{b_n}{a_n} \right)=
\left( \0{-r(y)}{2 y r'(y)} \right)_{y=y_{\rm min}} =\014\,.
\eeq
The first equality sign holds if the global minimum of the function
$y(1+r)$, attained at $y_{\rm min}$, is {\it local} in momentum space,
that is, not extending over an entire region around $y_{\rm min}$.
The second equality sign holds, if, in addition, the regulator
maximises \Eq{C}. To confirm \Eq{R}, it is useful to employ
$0=1+r+yr'$, which holds for optimised regulators at $y=y_{\rm min}$.
We conclude that higher order corrections due to $\eta$ are suppressed
for optimised regulators by an additional factor of (at least) $1/4$.
\step

Let us consider two explicit examples. For the quartic regulator and
any $d$, we find
\beq 
\label{power-eta}
\ell_{\rm quart}(\omega,\eta)
=\left(1-\0{\eta}{4}\right)\ell_{\rm quart}(\omega)\,.  \eeq
We conclude from \Eq{power-eta} that $b_n/a_n=\s014$ for {\it all}
$n$.  For the optimised regulator \Eq{ropt}, the minimum of the
function $y(1+r_{\rm opt}(y))$ extends over the entire interval $y\in
[0,1]$.  Therefore, the reasoning which has lead to \Eq{R} is not
applicable for \Eq{ropt}.  An explicit computation shows that
\cite{Litim:2001up}
\beq 
\label{opt-eta}
\ell_{\rm opt}(\omega,\eta)
=\left(1-\0{\eta}{d+2}\right)\ell_{\rm opt}(\omega)\,.  \eeq
We emphasize that the back-coupling of the anomalous dimension is {\it
  dimensionally} suppressed in \Eq{opt-eta} by a factor $1/(d+2)$, or
$\s015$ in $d=3$ dimensions. Actually, $b_n/a_n=1/(d+2)$ for {\it all}
$n$. Hence, of all optimised regulators, the regulator $r_{\rm opt}$
is singled out due to the {\it additional} suppression of $\eta$
corrections by a factor $4/(d+2)$ in comparison to generic optimised
flows.  The back coupling is also consistently smaller than for the
quartic regulator. Hence, we expect for optimised regulators that
$|\nu_1({\rm RS})/\nu_0({\rm RS})|$ is smallest for $r_{\rm opt}$ in
\eq{ropt}.

\begin{figure}[t]
\begin{center}
\unitlength0.001\hsize
\begin{picture}(700,600)
\put(110,430){
\begin{tabular}{ll}
Sharp&${}^{\multiput(0,0)(20,0){4}{\line(10,0){10}}} $
\\[-.7ex] 
Quart&${}^{\multiput(0,0)(20,0){3}{\put(0,0){\line(10,0){10}}
\put(14,0){\line(2,0){2}}}\put(60,0){\line(10,0){10}}}${}
\\[-.7ex] 
Opt&$  {}^{\put(0,0){\line(70,0){70}}}${}
\end{tabular}}
\put(300,70){{\large $N$}}
\put(515,110){$\infty$}
\put(120,500){ \fbox{{\huge $\nu$}{\large ${}_{\rm ERG}$}}}
\hskip.04\hsize
\psfig{file=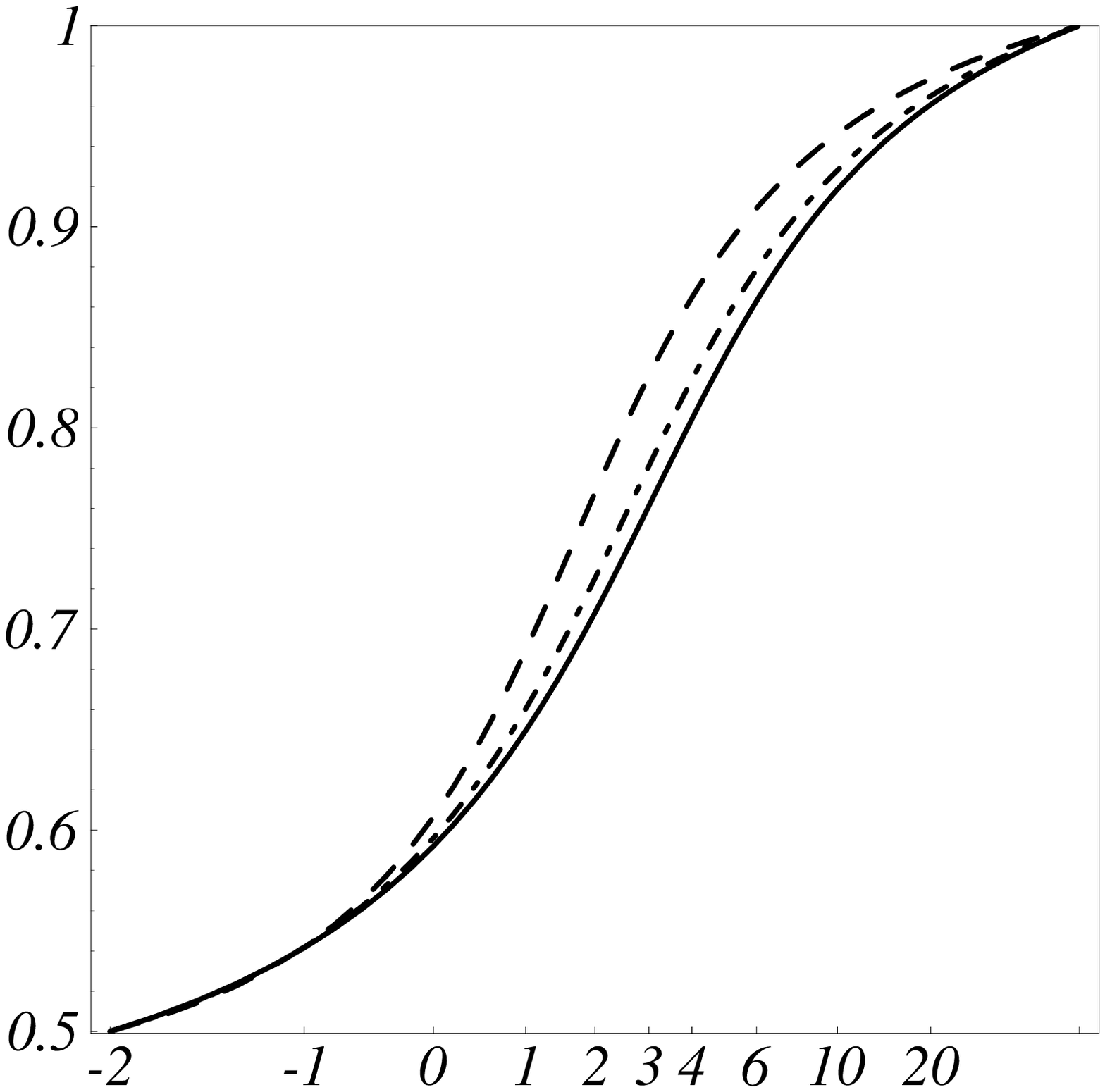,width=500\unitlength}
\end{picture}
\vskip-1.cm 
\begin{minipage}{.9\hsize}
  {\small {\bf Figure 1:} The critical exponent $\nu_{{}_{\rm ERG}}$ for
    various regulators. The $N$-axis has been squeezed as $N\to
    (N+2)/(N+6)$ for display purposes.}
\end{minipage} 
\end{center}
\end{figure}

\noindent 
{\bf 6. Critical exponents}\\[-1ex]

Starting with \Eq{GeneralFlow}, we have computed the critical
exponents $\nu$ and $\omega$ at the Wilson-Fisher (WF) fixed point.
For the numerics, we found it more convenient to use the flow for $u'$
instead of \Eq{GeneralFlow}.  In $d=3$ dimensions and for arbitrary
$N$, the flow equation \eq{GeneralFlow} has two fixed points
$u'_\star$ with $\partial_t u'_\star=0$: the trivial or Gau\ss ian
one, $u'_\star\equiv 0$, and the non-trivial WF fixed point with
$u'_\star \neq 0$.  Small perturbations about the WF fixed point have
a discrete spectrum of eigenvalues. The negative eigenvalue
corresponds to the unstable direction, and its negative inverse is
given by the critical exponent $\nu$. The exponent $\omega$ denotes
the smallest irrelevant (and hence positive) eigenvalue.  To leading
order in the derivative expansion, the anomalous dimension is
$\eta\equiv 0$.

\begin{figure}[t]
\begin{center}
\unitlength0.001\hsize
\begin{picture}(700,600)
\put(110,430){
\begin{tabular}{ll}
Sharp&${}^{\multiput(0,0)(20,0){4}{\line(10,0){10}}} $\\[-.7ex] 
Quart&${}^{\multiput(0,0)(20,0){3}{\put(0,0){\line(10,0){10}}
\put(14,0){\line(2,0){2}}}\put(60,0){\line(10,0){10}}}${}\\[-.7ex] 
Opt&$  {}^{\put(0,0){\line(70,0){70}}}${}
\end{tabular}}
\put(300,70){ {\large $N$}}
\put(515,110){$\infty$}
\put(120,500){ \fbox{{\huge $\omega$}{\large ${}_{\rm ERG}$}}}
\hskip.04\hsize
\psfig{file=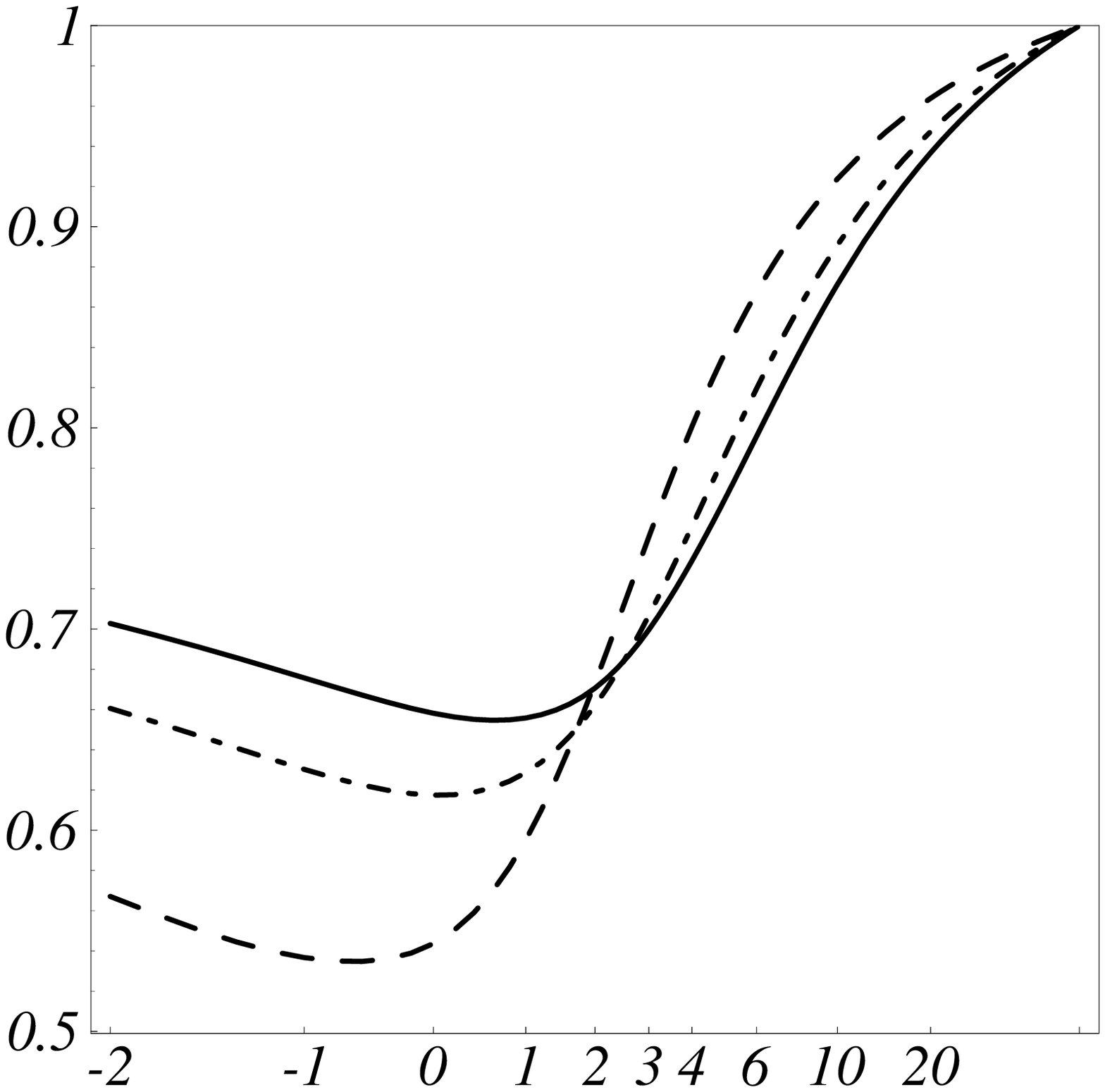,width=500\unitlength}
\end{picture}
\vskip-1.cm 
\begin{minipage}{.9\hsize}
  \small {\bf Figure 2:} The smallest irrelevant eigenvalue at
  criticality, $\omega_{{}_{\rm ERG}}$. The $N$-axis has been squeezed
  as $N\to (N+2)/(N+6)$ for display purposes.
\end{minipage} 
\end{center}
\end{figure}

Fig.~1 shows our results for the exponent $\nu(N)$ for $-2\le N\le
\infty$, and for the three classes of regulators $r_{\rm opt}$,
$r_{\rm quart}$ and $r_{\rm sharp}$. For the sharp cut off, our
results agree with the findings
of~\cite{Hasenfratz:1986dm,Comellas:1997tf}. For all regulators, the
critical exponent $\nu(N)$ is a monotonically increasing function with
$N$. For fixed $N$, the critical exponent $\nu$ depends on the RS.
Notice that the curves of $r_{\rm opt}$ and $r_{\rm quart}$ are
essentially on top of each other for $N\le -1$.  The results of
Ref.~\cite{Liao:2000sh} are also worth mentioning.  There, the authors
computed the critical exponent $\nu$ for $N=1$ and the classes of
regulators \Es{rpower}, \eq{rexp} and \eq{rmix} for all $b\ge 1$,
finding $\nu_{\rm sharp}\ge\nu_b>\nu_{\rm opt}$.  More generally, the
critical exponents are bounded from below $\nu_{{}_{\rm ERG}}\ge
\nu_{\rm opt}$ to leading order in the derivative expansion
\cite{Litim:CritExp}, and for $N\gta -1$, the smallest value is
attained through $r_{\rm opt}$.  For $N\lta -1$, we note that the
line for $r_{\rm sharp}$ is below those for $r_{\rm opt}$ and $r_{\rm
  quart}$.  In Fig.~1, however, this small effect is barely visible.
\step

Fig.~2 shows the first irrelevant critical exponent $\omega(N)$ for
$-2\le N\le \infty$. In contrast to $\nu$, it is no longer a monotonic
function of $N$. As a function of the regularisation, the turning
point $\partial\omega/\partial N=0$ moves from $N\approx 1$ ($r_{\rm
  opt}$) via $N\approx 0$ ($r_{\rm quart}$) to $N\approx-1$ ($r_{\rm
  sharp}$).  We emphasize that all subleading eigenvalues for
different RS join in the limit $N=\infty$, in consistency with the
known result \Eq{omega-largeN}. However, this holds not true for
$N=-2$, where $\omega$ found to be non-universal.

\begin{figure}[t]
\begin{center}
  \unitlength0.001\hsize
\begin{picture}(700,600)
\put(370,500){
\begin{tabular}{ll}
Sharp&${}^{\multiput(0,0)(20,0){4}{\line(10,0){10}}} $\\[-.7ex] 
Quart&${}^{\multiput(0,0)(20,0){3}{\put(0,0){\line(10,0){10}}
\put(14,0){\line(2,0){2}}}\put(60,0){\line(10,0){10}}}${}\\[-.7ex] 
Opt&$  {}^{\put(0,0){\line(70,0){70}}}${}
\end{tabular}}
\put(515,120){$\infty$}
\put(300,80){ {\large $N$}}
\put(390,400){ \fbox{{\large $\displaystyle \0{\nu_{{}_{\rm \tiny ERG}}}{\nu_{\rm opt}}-1$}}}
\hskip.04\hsize
\psfig{file=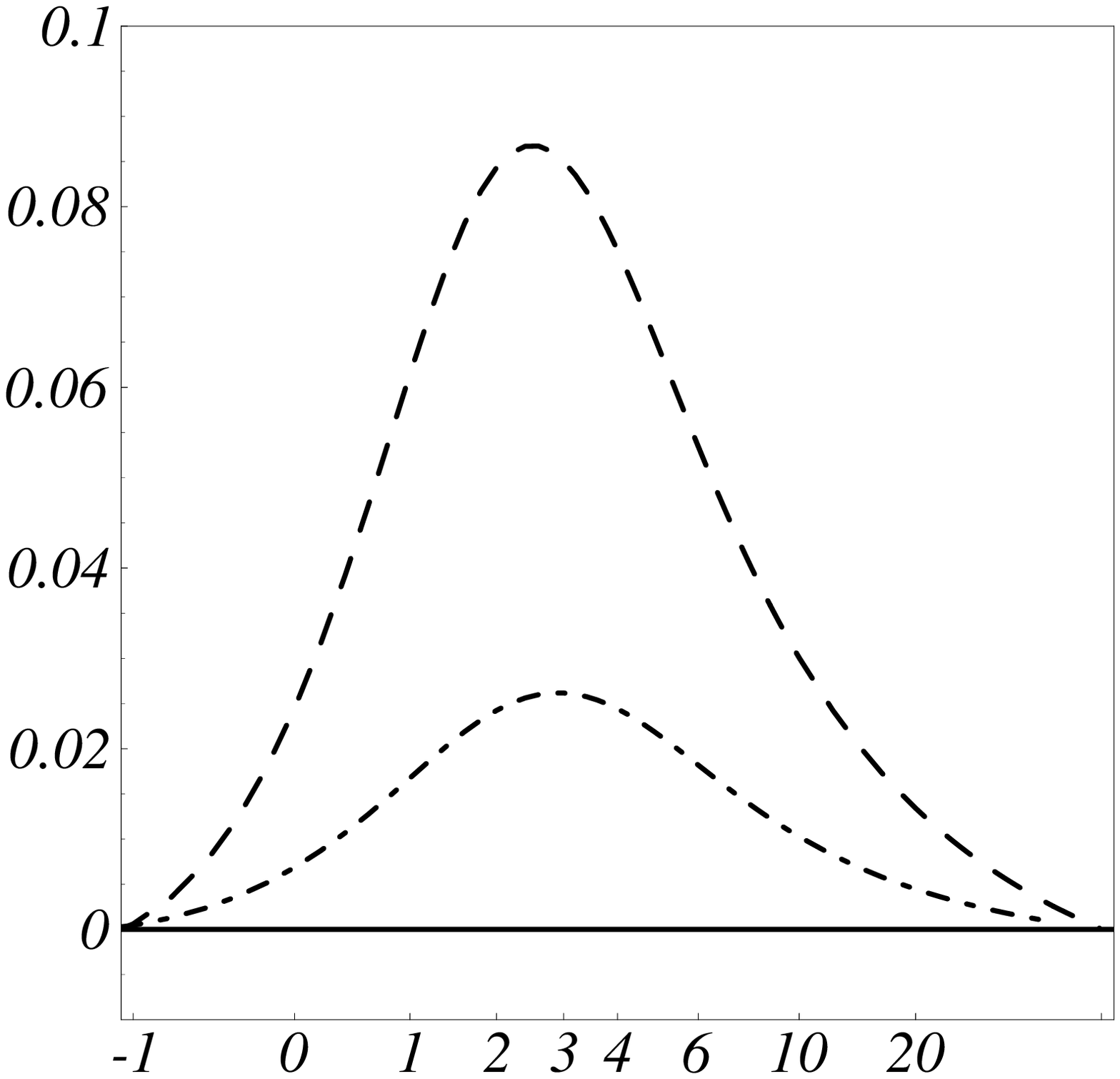,width=500\unitlength}
\end{picture}
\vskip-1.cm 
\begin{minipage}{.9\hsize}
  {\small {\bf Figure 3:} The relative improvement of the critical
    exponents $\nu_{{}_{\rm \tiny ERG}}$ for various regulators in comparison
    with $\nu_{\rm opt}$. }
\end{minipage} 
\end{center}
\end{figure}

As a good measure for the scheme dependence, at any fixed $N$, we
consider the range over which $\nu$(RS) and $\omega$(RS) vary as
functions of the RS. Such a comparison is sensible, because a variety
of qualitatively different regulators are covered within the
boundaries set by the sharp and the smooth optimised cutoff. In
Fig.~3, we have displayed the ratio ${\nu_{{}_{\rm ERG}}}/{\nu_{\rm
    opt}}-1$ for the sharp cut off (dashed line) and the quartic
(dashed-dotted line) regulator. For $\nu$, the boundaries are set by
regulators with the largest (smallest) gap $C$, and hence by $r_{\rm
  sharp}$ and $r_{\rm opt}$ with $C_{\rm sharp}/C_{\rm opt}=\s012$.
From Fig.~3, we conclude that the scheme dependence vanishes for
$N=\infty$ and $N=-2$, consistent with the known results
\Es{nu-largeN} and \eq{nu-2}. In turn, the scheme dependence is
largest around $N_{\rm max}\approx 2-3$.  This suggests that a
perturbative expansion around $N=\infty$ ($N=-2$) may be feasible for
$N>N_{\rm max}$ ($N<N_{\rm max}$).  The scheme dependence of $\omega$
(Fig.~2) turns out to be largest at $N=-2$, while it vanishes at
$N=\infty$.  \step

Based only on Fig.~3, it is not possible to decide which regulator
would lead to a good estimate of the physical value within the given
approximation. At this point, we take advantage of the reasoning
presented in sections 4 and 5, where it has been argued that an
optimised regulator leads to better convergence and stability
properties of the flow, and to smaller higher order corrections.
Therefore, $r_{\rm opt}$ provides a prefered choice. In this light,
the comparison in Fig.~3 gives a quantitative idea on the improvement
implied by an optimised choice for the regulator.  With respect to the
sharp cut off, the optimised regulator leads to a decrease of
$\nu_{{}_{\rm ERG}}$ up to $10\%$ (or $3-10\%$ for $N\in [0,4]$; the
maximum is attained around $N\approx 2-3$). In comparison to the
quartic regulator, the optimised regulator $r_{\rm opt}$ leads to a
decrease of up to $3\%$, with the maximum around $N\approx 3$.  For
reasons given earlier, the relative decrease is smaller compared to
the sharp cut off.  The values for $\nu_{\rm opt}$ are indeed closest
to the physical values, to leading order in the derivative expansion.
\\[3ex]

\noindent 
{\bf 7. Comparison}\\[-1ex]

Results by other renormalisation group methods, and experiment, have
been discussed in the literature. We compare the findings for
$\nu_{{}_{\rm ERG}}$ and $\omega_{{}_{\rm ERG}}$ with ERGs for other
regulators
\cite{Hasenfratz:1986dm,Morris:1994ie,Comellas:1997tf,Morris:1998xj,Liao:2000sh,VonGersdorff:2000kp},
with the Polchinski RG \cite{Polchinski,Ball:1995ji,Comellas:1998ep},
with a proper time RG \cite{Litim:2001hk,Mazza:2001bp}, all to leading
and subleading order in the derivative expansion, and with
experimental data, data from Monte Carlo simulations
\cite{Pelissetto:2000ek} and results from other field theoretical
methods \cite{Zinn-Justin:2001bf}. In Fig.~4 (Fig.~5), the critical
index $\nu$ (subleading eigenvalue $\omega$) is displayed as predicted
by various methods. For $\nu$, the range of data points is taken from
\cite{Pelissetto:2000ek}. For $N=0, 3$ and $4$, the bounday values
stem from field theoretical predictions or MC simulations. For $N=1$
and $2$, the boundary values stem from experimental data. The bounds
on field theoretical methods, reviewed in \cite{Zinn-Justin:2001bf},
are much tighter. They agree surprisingly well with the findings of
\cite{Litim:2001hk} (within less than $1\%$ for $\nu$ and around $5\%$
for $\omega$, for all $N=0,\cdots,4$), and are represented by the
short-dashed line in Figs.~4 and 5.\step

To order ${\cal O}(\partial^0)$, results for $\nu$ (Fig.~4) and
$\omega$ (Fig.~5) are given for the sharp (medium dashed line), the
quartic (dashed-dotted line) and the optimised (full line) regulator.
In \cite{Litim:2001fd}, it has been shown that the critical exponents
from the Polchinski RG, at the present order, are equivalent to those
from the optimised ERG. All ERG results for $\nu_{{}_{\rm ERG}}$ are
systematically too large.  Changing the ERG regulator from sharp via
quartic to the optimised one, we notice in Fig.~4 that the curves bend
down towards the values favoured by experiments and other methods. In
particular, the results for \Eq{ropt} are closest to the physical
values. This behaviour is fully consistent with the picture derived
above: we expect that higher order corrections for $r_{\rm opt}$ are
smaller than those for $r_{\rm sharp}$ and $r_{\rm quartic}$. In
Fig.~5, we notice that the $\omega_{{}_{\rm ERG}}$ curves for the
sharp, the quartic and the optimised regulator bend towards the values
prefered by other methods. The $N$ dependence of $\omega_{{}_{\rm
    ERG}}$ matches the $N$ dependence of the data only for small $N$.
Still, the numerical agreement is acceptable, bearing in mind that
$\omega_{{}_{\rm ERG}}$ is a subleading exponent like $\eta$.  
\step

To order ${\cal O}(\partial^2)$, the exponential regulator
\cite{VonGersdorff:2000kp} leads to good predictions for $\nu$, for
all $N$ considered (Fig.~4). No results for $\omega$ have been
reported. Within the Polchinski RG, results for $N=1$ have been given
in \cite{Ball:1995ji,Comellas:1998ep}. To order ${\cal
  O}(\partial^2)$, the results depend on two scheme-dependent
parameters, which are determined by matching the anomalous dimension
with the known experimental value, and by a minimum sensitivity
condition. Then, the results for $\nu$ and $\omega$ agree well with
other methods.

\begin{figure}[t]
\begin{center}
\unitlength0.001\hsize
\begin{picture}(600,700)
\put(110,470){
\begin{tabular}{ll}
${\cal O}(\partial^0)$&\\
Sharp&${}^{\multiput(0,0)(20,0){4}{\line(10,0){10}}} $\\
Quart&${}^{\multiput(0,0)(20,0){3}{\put(0,0){\line(10,0){10}}
           \put(14,0){\line(2,0){2}}}\put(60,0){\line(10,0){10}}}${}\\
Opt  &$  {}^{\put(0,0){\line(70,0){70}}}${}\\ 
PT   &$  {}^{\multiput(0,0)(10,0){8}{\line(5,0){5}}}$\\[2ex]
${\cal O}(\partial^2)$&\\
Exp  &$\ \bullet$ \\
Quart&$\ \bullet${\footnotesize \it a}\\
PT   &$\ \bullet${\footnotesize \it b}\\
PRG  &$\ \bullet${\footnotesize \it c}
\end{tabular}
}
\put(257,157){${}_a$}
\put(257,170){${}_{b,c}$}
\put(328,217){${}_a$}
\put(400,390){${}_a$}
\put(468,520){${}_a$}
\put(300,-20){ {\large $N$}}
\put(450,100){{{\Huge $\nu$}}}
\hskip.04\hsize
\psfig{file=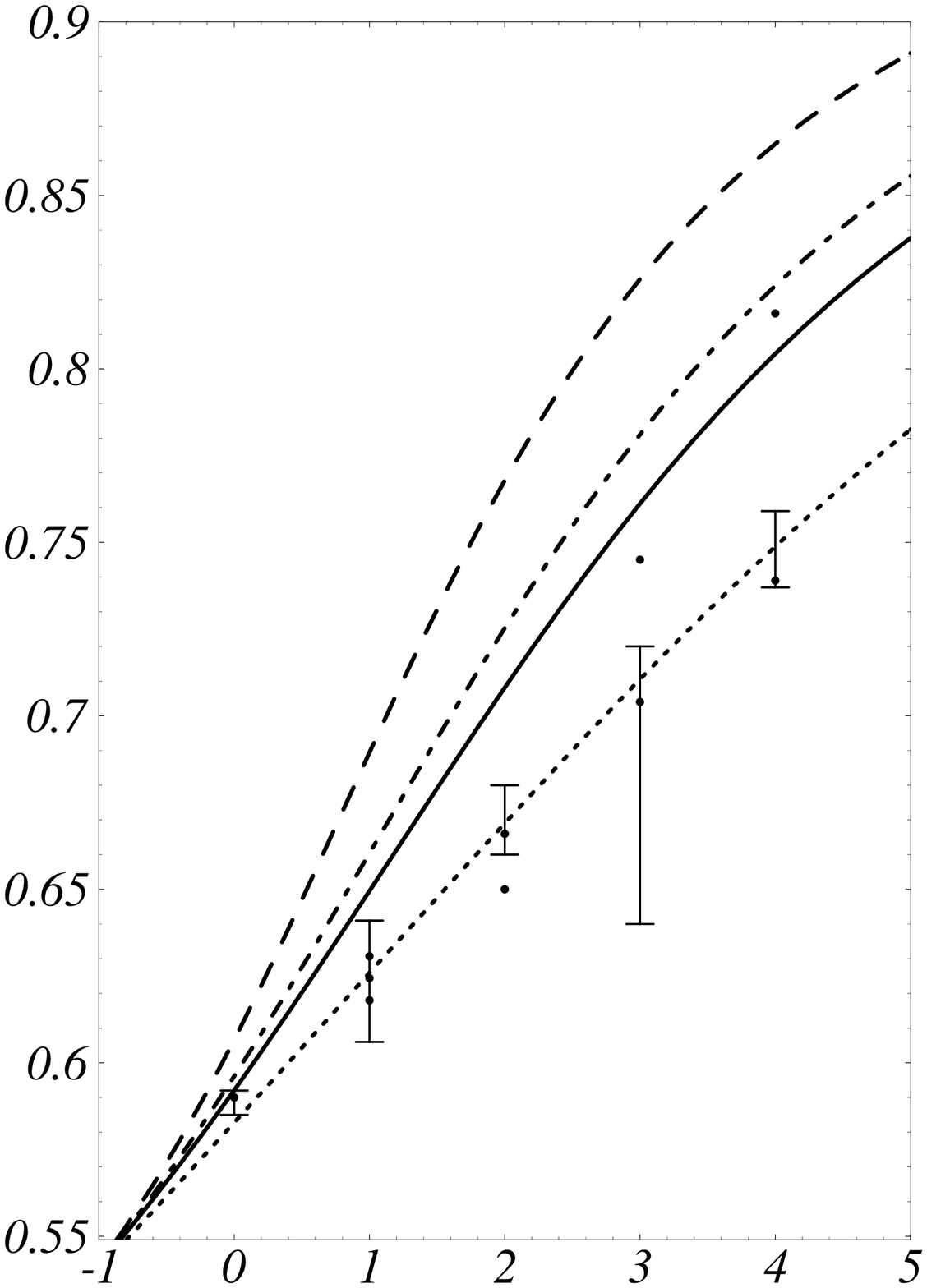,width=500\unitlength}
\end{picture}
\vskip1.cm 
\begin{minipage}{.9\hsize}
  \small {\bf Figure 4:} The critical exponent $\nu$; comparison of
  ERG results at ${\cal O}(\partial^0)$ (Sharp, Quartic, Opt) with a
  specific proper time RG (PT) \cite{Litim:2001hk}, with ERG results
  to order ${\cal O}(\partial^2)$ (Quartic, Exp)
  \cite{Morris:1994ie,Morris:1998xj,VonGersdorff:2000kp}, with results
  from the Polchinski RG (PRG) \cite{Ball:1995ji,Comellas:1998ep} and
  a proper time RG (PT) to order ${\cal O}(\partial^2)$
  \cite{Mazza:2001bp}, and with results from other methods (MC
  simulations, high temperature expansions, other field theoretical
  methods, experiments) \cite{Pelissetto:2000ek}.
\end{minipage} 
\end{center}
\end{figure}

For the quartic regulator \cite{Morris:1994ie,Morris:1998xj}, the
results for $\nu$ agree less well with the available data, except for
$N=1$.  For $N=4$, the ${\cal O}(\partial^2)$ result for the quartic
regulator is already worse than the ${\cal O}(\partial^0)$ result
based on \Eq{ropt}. The results for $\omega$ (Fig.~5) are even more
sensitive and show a strong $N$ dependence. The findings of
Ref.~\cite{Morris:1998xj} suggest that the derivative expansion in
$d=3$ dimensions with the quartic regulator converges less well for
intermediate $N>1$ (see also the related comments in
Ref.~\cite{Morris:1998xj}).  Supposedly, this problematic behaviour is
due to the weak UV behaviour of the quartic regulator in $3d$. In the
UV limit, $\partial_t R_{\rm quart}$ in \Eq{ERG} vanishes only
{polynomially}, but not {exponentially}.  This corresponds to an
insufficiency in the integrating out of UV modes for \Eq{rpower}, and
may spoil the convergence.  The optimisation ideas discussed in
section 4 are essentially sensitive to the IR behaviour of the
regulator, and cannot detect insufficiencies coming from the UV
behaviour.

\begin{figure}[t]
\begin{center}
\unitlength0.001\hsize
\begin{picture}(600,700)
\put(245,445){${}_c$}
\put(250,535){${}_b$}
\put(250,580){${}_a$}
\put(318,105){${}_a$}
\put(395,55){${}_a$}
\put(465,140){${}_a$}
\put(350,250){
\begin{tabular}{ll}
${\cal O}(\partial^0)$&\\
Sharp &${}^{\multiput(0,0)(20,0){4}{\line(10,0){10}}}  $ \\
Quart &${}^{\multiput(0,0)(20,0){3}{\put(0,0){\line(10,0){10}}
\put(14,0){\line(2,0){2}}}\put(60,0){\line(10,0){10}}}$ \\
Opt   &${}^{\put(0,0){\line(70,0){70}}}${}\\ 
PT&$  {}^{\multiput(0,0)(10,0){8}{\line(5,0){5}}}$ {}
\end{tabular}}
\put(380,590){
\begin{tabular}{ll}
${\cal O}(\partial^2)$&\\
Quart&$\ \ \bullet${\footnotesize \it a}\\
PT&$\ \ \bullet${\footnotesize \it b}\\
PRG&$\ \ \bullet${\footnotesize \it c}
\end{tabular}}
\put(290,-25){ {\large $N$}}
\put(150,100){ {{\Huge $\omega$}}}
\hskip.04\hsize
\psfig{file=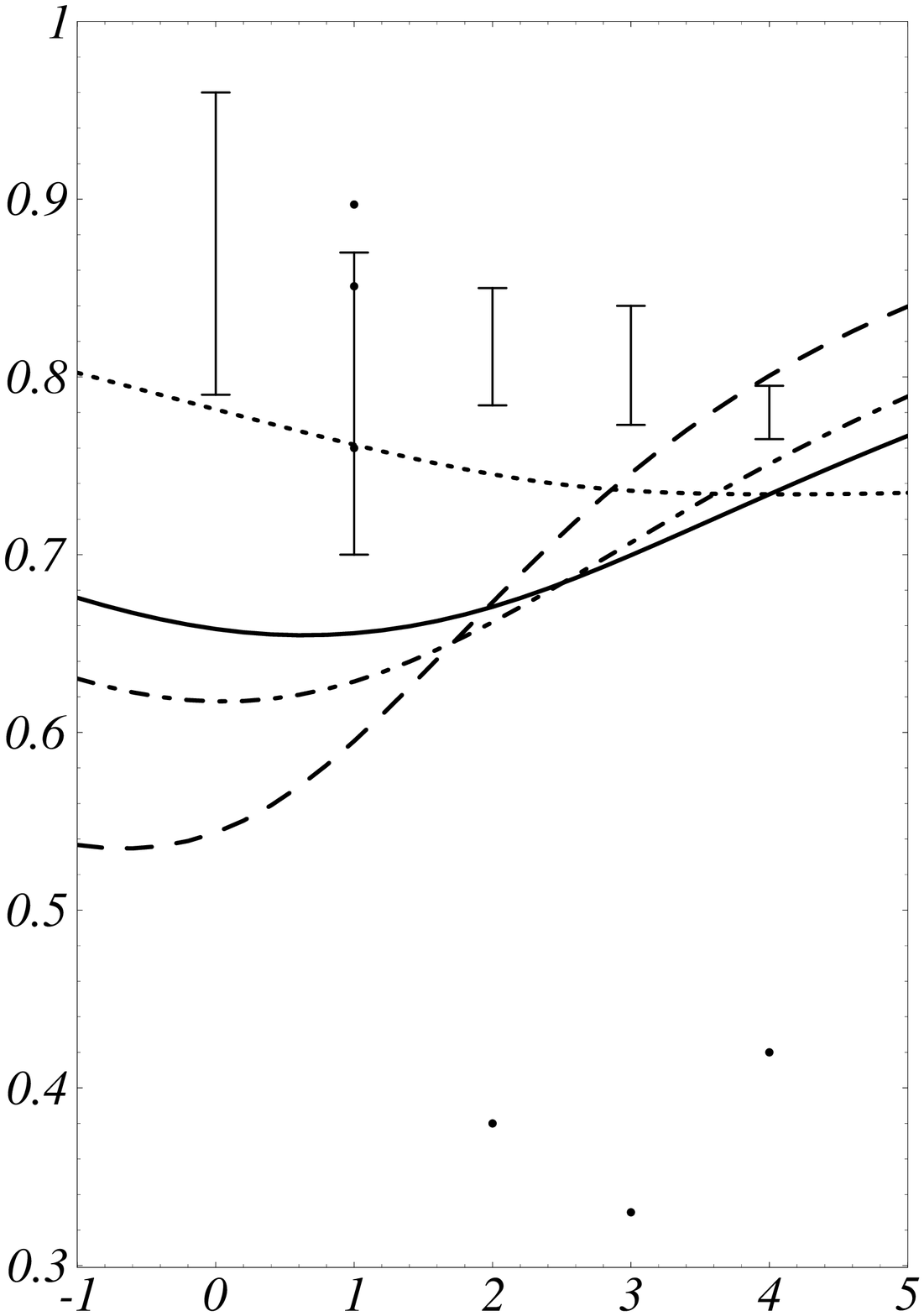,width=500\unitlength}
\end{picture}
\vskip1.cm 
\begin{minipage}{.9\hsize}
  {\small {\bf Figure 5:} The smallest subleading eigenvalue $\omega$;
    comparison of ERG results to order ${\cal O}(\partial^0)$ (Sharp,
    Quartic, Opt) with a specific proper time RG (PT) to order ${\cal
      O}(\partial^0)$ \cite{Litim:2001hk}, and predictions by other
    methods \cite{Pelissetto:2000ek}. To order ${\cal O}(\partial^2)$,
    the ERG results (quartic regulator) are from \cite{Morris:1994ie},
    the Polchinski RG (PRG) result is from
    \cite{Ball:1995ji,Comellas:1998ep}, and the PT result from
    \cite{Mazza:2001bp}.}
\end{minipage} 
\end{center}
\end{figure}

Finally, we turn to a proper time (PT) renormalisation group
\cite{Liao:1996fp} for a specific operator cutoff discussed in
Ref.~\cite{Litim:2001hk}. In contrast to the ERG flow \eq{ERG}, a PT
flow is {\it not} an exact one (see Ref.~\cite{Litim:2001hk} for a
comparison of ERG and PT flows).  To leading order in the derivative
expansion, only a subset of PT flows can be mapped onto ERG flows.
Here, we consider a specific PT flow which cannot be mapped on a
corresponding ERG flow, namely
\beq\label{PT}
\partial_t\Gamma_k=\Tr \,\exp-\Gamma_k^{(2)}/k^2\,.
\eeq
Currently, it is not known what approximation to an exact flow it
describes \cite{Litim:2001hk}. To leading order in the derivative
expansion, \Eq{PT} turns into \Eq{GeneralFlow} with $\ell_{\rm
  PT}(\omega)=\Gamma(\s0d2)\exp(-\omega)$
\cite{Litim:2001hk,Mazza:2001bp}.  Next, we apply the reasoning of
section 4 to $\ell_{\rm PT}(\omega)$. Expanding the function
$\ell_{\rm PT}(\omega)$ as in \Eq{ell-w}, the effective radius of
convergence \Eq{C} of the flow \Eq{PT} is trivially found to be
$C_{\rm eff}=\infty$. Notice that $C_{\rm eff}$ is read off from
$\ell_{\rm PT}(\omega)$, and not from a regularised propagator as in
\Eq{Cexpl}. The reason is simple. There exists no map which brings
\eq{PT} into the form \eq{ERG}, not even to leading order in the
derivative expansion \cite{Litim:2001hk}. The structure of the exact
RG --- with normalised regulators as defined in \Eq{r} and $C$ given
by \Eq{Cexpl} --- implies $C_{\rm ERG}<\infty$. Based on $C_{\rm
  eff}>C_{\rm ERG}$, we expect that the derivative expansion for
\Eq{PT} converges rapidly.\step

Critical exponents have been computed from \Eq{PT} in
Refs.~\cite{Litim:2001hk,Mazza:2001bp} (see also
Ref.~\cite{Bonanno:2000yp}). To order ${\cal O}(\partial^0)$, the PT
results for $\nu$ $(\omega)$ are displayed in Fig.~4 (Fig.~5) by a
short dashed line. The PT results for $\nu(N)$ agree within a percent
with those from other field theoretical methods
\cite{Zinn-Justin:2001bf}.  The $N$ dependence of $\omega(N)$ is
reproduced within $5\%$.  Notice that the ${\cal O}(\partial^0)$ PT
results nearly coincide with the ${\cal O}(\partial^2)$ ERG result for
the exponential regulator.  Furthermore, the leading order PT result
for $N=1$ agrees with the second order result of the Polchinski RG
\cite{Ball:1995ji,Comellas:1998ep}. To order ${\cal O}(\partial^2)$,
for $N=1$, the PT flow induces a negligible correction to $\nu$, and a
$10\%$ increase to $\omega$ \cite{Mazza:2001bp}. While these results
confirm our reasoning based on the arguments explained in sections 4
and 5, the numerical agreement with results obtained by other methods,
or experiment, remains unclear.  Currently, it is not understood {why}
the derivative expansion of \Eq{PT} should converge towards the {\it
  physical} scaling solution, bearing in mind that \Eq{PT} is not an
exact flow \cite{Litim:2001hk}.  An answer to this question, however,
is outside the range of the present study.
\\[3ex]

\noindent 
{\bf 8. Discussion}\\[-1ex]

We have studied the convergence of the derivative expansion. An
adequate choice of the IR regulator turned out to be most vital for a
good convergence. This is linked to the spurious regulator dependence
of approximated flows, a dependence which would vanish for the
integrated full flow. Indeed, changing the IR regulator within an
approximated flow can be seen as a slight reorganisation of the
derivative expansion. At a given order in the expansion, an optimised
regulator effectively leads to the incorporation of contributions,
which for other regulators would have appeared only to higher order.
This is why certain regulators lead to better results already at lower
orders. This interplay is used to {\it improve} the convergence of the
derivative expansion.\step

We have applied these observations to the exact renormalisation group,
to leading order in the derivative expansion and for subleading
corrections proportional to the anomalous dimension. The leading order
results for critical exponents of 3d $O(N)$ symmetric scalar theories
are improved significantly for specific optimised regulators (Fig.~3).
Furthermore, the back coupling of the anomalous dimension is reduced
by a factor of $1/4$ for generic optimised regulators, and by
$1/(d+2)$ for the specific regulator \Eq{ropt}.  This provides an
additional explanation for why higher order corrections remain small.
For these reasons --- in combination with the exactness of the flow
--- we expect that the corresponding critical exponents are closer to
those of the physical theory.  \step

This picture has explicitly been confirmed by establishing a link
between the radius of convergence (a.k.a.~the ``gap'') and the
proximity of the corresponding value for $\nu_{{}_{\rm ERG}}$ to the
physical value $\nu_{\rm phys}$. The line of reasoning is entirely
based on the structure of the flow \eq{ERG}, which contains all the
universal information relevant at a scaling solution.  In principle,
our analysis applies as well for other physical observables which can
be computed for small field amplitudes.  As a side result, we found
that the subleading critical exponents for $N=-2$ are non-universal,
in contrast to the case $N=\infty$.  \step

For the proper-time renormalisation group, a reasoning analogous to
the one in section 4 has been applied. We have considered the radius
of convergence for amplitude expansions of the specific PT flow
\eq{PT}, which is found to be larger than for exact RG flows. This
qualitative difference between exact and PT flows is linked to their
fundamental inequivalence, even to this order of the approximation
\cite{Litim:2001hk}. Still, the large effective radius of convergence
explains why the derivative expansion for the PT flow \Eq{PT} should
have even better convergence properties. While this explains the
convergence behaviour found in Ref.~\cite{Mazza:2001bp} for $N=1$, it
remains to be clarified whether it converges towards the physical
theory.\step

It would be interesting to apply these considerations to other
theories, like scalar QED
\cite{Litim:1997jd,ScalarQED-N,ScalarQED,Freire:2001sx}, or quantum
Einstein gravity \cite{QuantumGravity1}, where recent
investigations showed strong evidence for the existence of a
non-trivial UV fixed point even in $d=4$ dimensions. In contrast to
the scalar theories studied here, the stability matrix at criticality
discussed in \cite{QuantumGravity2} has complex
eigenvalues for various classes of regulators, and a minimum
sensitivity condition seems not to be applicable.  Within the
approximations employed in \cite{QuantumGravity2}, we
expect that the optimised flows of \cite{Litim:2001up} should lead to
a good prediction of the fixed point.
\\[2ex]
\noindent 
{\it Acknowledgements:} This work has been supported by a Marie-Curie
fellowship under EC contract no.~HPMF-CT-1999-00404.
\\[3ex]

\noindent 
{\bf References}\\[-1ex]

\end{document}